\documentclass[fleqn,twoside]{article}
\usepackage{espcrc2}

\usepackage{graphicx}
\usepackage[figuresright]{rotating}

\newcommand{\AmS}{{\protect\the\textfont2
  A\kern-.1667em\lower.5ex\hbox{M}\kern-.125emS}}
\newcommand{\beqn}{\begin{eqnarray}}
\newcommand{\eeqn}{\end{eqnarray}}
\newcommand{\be}{\begin{equation}}
\newcommand{\ee}{\end{equation}}

\def\beq{\begin{equation}}
\def\be{\begin{equation}}
\def\beqn{\begin{eqnarray}}
\def\ee{\end{equation}}
\def\eeq{\end{equation}}
\def\eeqn{\end{eqnarray}}

\def\PL{Phys. Lett.}
\def\PRL{Phys. Rev. Lett.}

\def\PR{Phys. Rev.}

\def\beq{\begin{equation}}
\def\eeq{\end{equation}}

\def\NCA{\em Nuovo Cimento}
\def\NIM{\em Nucl. Instrum. Methods}
\def\NIMA{{\em Nucl. Instrum. Methods} A}
\def\NPA{{\em Nucl. Phys.} A}
\def\NPB{{\em Nucl. Phys.} B}
\def\PLB{{\em Phys. Lett.}  B}
\def\PRL{\em Phys. Rev. Lett.}
\def\PRD{{\em Phys. Rev.} D}
\def\ZPC{{\em Z. Phys.} C}

\def\s1{$s_{\alpha}$}
\def\s2{$s_{\gamma}$}
\def\s3{$s_{\delta}$}
\def\c1{$c_{\alpha}$}
\def\c2{$c_{\gamma}$}
\def\c3{$c_{\delta}$}
\def\s{Stueckelberg~}

\newcommand{\mathsym}[1]{{}}

\def\PL{Phys. Lett.}
\def\PRL{Phys. Rev. Lett.}

\def\PR{Phys. Rev.}

\def\beq{\begin{equation}}
\def\eeq{\end{equation}}
\def\beqn{\begin{eqnarray}}
\def\eeqn{\end{eqnarray}}

\def\NCA{\em Nuovo Cimento}
\def\NIM{\em Nucl. Instrum. Methods}
\def\NIMA{{\em Nucl. Instrum. Methods} A}
\def\NPA{{\em Nucl. Phys.} A}
\def\NPB{{\em Nucl. Phys.} B}
\def\PLB{{\em Phys. Lett.}  B}
\def\PRL{\em Phys. Rev. Lett.}
\def\PRD{{\em Phys. Rev.} D}
\def\ZPC{{\em Z. Phys.} C}

\def\s1{$s_{\alpha}$}
\def\s2{$s_{\gamma}$}
\def\s3{$s_{\delta}$}
\def\c1{$c_{\alpha}$}
\def\c2{$c_{\gamma}$}
\def\c3{$c_{\delta}$}
\def\s{Stueckelberg~}

\def \tb{\widetilde{t}_2}

\def\beq{\begin{equation}}
\def\be{\begin{equation}}
\def\beqn{\begin{eqnarray}}
\def\ee{\end{equation}}
\def\eeq{\end{equation}}
\def\eeqn{\end{eqnarray}}

\def\beq{\begin{equation}}
\def\be{\begin{equation}}
\def\beqn{\begin{eqnarray}}
\def\ee{\end{equation}}
\def\eeq{\end{equation}}
\def\eeqn{\end{eqnarray}}

\def \tb{\widetilde{t}_2}

\def \tb{\widetilde{t}_2}

\def\.4{\vspace{-.5cm}}
\def\br{\left(\begin{array}{c}}
\def\er{\end{array}\right)}

\def\c{GNLSP$_{\rm C}~$}

\usepackage{amssymb,graphics,graphicx, epsfig}
\usepackage{amsfonts}
\usepackage{hyperref}
\usepackage{multicol}
\usepackage{bm}
\usepackage{hyperref}
\title{On the Possible Observation of  Mirror Matter}

\author{Tarek Ibrahim
\address{Department of Physics, University of Alexandria, Egypt}
and Pran Nath 
\address{Department of Physics,  Northeastern University, Boston, MA 02115, USA}}

\begin{document}

\begin{abstract}
The possibility that mirror matter with masses in the several hundred 
GeV- TeV  range exists  is explored. Mirror matter appears quite naturally in many unified models
of particle interactions both in GUTs and in strings often in vector-like combinations. 
Some of  these vector-like multiplets could escape 
acquiring super heavy masses and remain light down to the low energies where they 
acquire vector-like masses of electroweak size.
It is found that a very small mixing of the vector-like multiplets with  MSSM matter (specifically
with the third generation matter)  can produce very large contributions to the magnetic 
moment of the $\tau$ neutrino  by as much as several orders of magnitude putting this 
moment in the range of accessibility of improved experiment. 
Further, it is shown that if mirror matter exists it would lead to distinctive signatures at colliders and thus such
matter  can be  explored at the  LHC  energies with available luminosities.

\vspace{1pc}
\end{abstract}
\maketitle

\def\NCA{\em Nuovo Cimento}
\def\NIM{\em Nucl. Instrum. Methods}
\def\NIMA{{\em Nucl. Instrum. Methods} A}
\def\NPA{{\em Nucl. Phys.} A}
\def\NPB{{\em Nucl. Phys.} B}
\def\PLB{{\em Phys. Lett.}  B}
\def\PRL{\em Phys. Rev. Lett.}
\def\PRD{{\em Phys. Rev.} D}
\def\ZPC{{\em Z. Phys.} C}
\def\rmuu{\gamma^{\mu}}
\def\rmud{\gamma_{\mu}}
\def\PL{{1-\gamma_5\over 2}}
\def\PR{{1+\gamma_5\over 2}}
\def\sinW2{\sin^2\theta_W}
\def\AEM{\alpha_{EM}}
\def\mul{M_{\tilde{u} L}^2}
\def\mur{M_{\tilde{u} R}^2}
\def\mdl{M_{\tilde{d} L}^2}
\def\mdr{M_{\tilde{d} R}^2}
\def\mz2{M_{z}^2}
\def\c2b{\cos 2\beta}
\def\au{A_u}
\def\ad{A_d}
\def\cob{\cot \beta}
\def\v#1{v_#1}
\def\tb{\tan\beta}
\def\epem{$e^+e^-$}
\def\KK{$K^0$-$\overline{K^0}$}
\def\wi{\omega_i}
\def\xj{\chi_j}
\def\Wmu{W_\mu}
\def\Wnu{W_\nu}
\def\m#1{{\tilde m}_#1}
\def\mH{m_H}
\def\mw#1{{\tilde m}_{\omega #1}}
\def\mx#1{{\tilde m}_{\chi^{0}_#1}}
\def\mc#1{{\tilde m}_{\chi^{+}_#1}}
\def\mwi{{\tilde m}_{\omega i}}
\def\mxi{{\tilde m}_{\chi^{0}_i}}
\def\mci{{\tilde m}_{\chi^{+}_i}}

\def\ch{{\tilde\chi^{+}_1}}
\def\c2{{\tilde\chi^{+}_2}}

\def\tt{{\tilde\theta}}

\def\tp{{\tilde\phi}}

\def\mz{M_z}
\def\sw{\sin\theta_W}
\def\cw{\cos\theta_W}
\def\cb{\cos\beta}
\def\sb{\sin\beta}
\def\rwi{r_{\omega i}}
\def\rxj{r_{\chi j}}
\def\rfp{r_f'}
\def\Kik{K_{ik}}
\def\Fq2{F_{2}(q^2)}
\def\f{\({\cal F}\)}
\def\d1{{\f(\tilde c;\tilde s;\tilde W)+ \f(\tilde c;\tilde \mu;\tilde W)}}
\def\tw{\tan\theta_W}
\def\sec2w{sec^2\theta_W}
\def\lsim{\ ^<\llap{$_\sim$}\ }
\def\gsim{\ ^>\llap{$_\sim$}\ }
\def\r2{\sqrt 2}
\def\beq{\begin{equation}}
\def\eeq{\end{equation}}
\def\beqn{\begin{eqnarray}}
\def\eeqn{\end{eqnarray}}
\def\rmuu{\gamma^{\mu}}
\def\rmud{\gamma_{\mu}}
\def\PL{{1-\gamma_5\over 2}}
\def\PR{{1+\gamma_5\over 2}}
\def\sinW2{\sin^2\theta_W}
\def\AEM{\alpha_{EM}}
\def\mul{M_{\tilde{u} L}^2}
\def\mur{M_{\tilde{u} R}^2}
\def\mdl{M_{\tilde{d} L}^2}
\def\mdr{M_{\tilde{d} R}^2}
\def\mz2{M_{z}^2}
\def\c2b{\cos 2\beta}
\def\au{A_u}         
\def\ad{A_d}
\def\cob{\cot \beta}
\def\v#1{v_#1}
\def\tb{\tan\beta}
\def\epem{$e^+e^-$}
\def\KK{$K^0$-$\bar{K^0}$}
\def\wi{\omega_i}
\def\xj{\chi_j}
\def\Wmu{W_\mu}
\def\Wnu{W_\nu}
\def\m#1{{\tilde m}_#1}
\def\mH{m_H}
\def\mw#1{{\tilde m}_{\omega #1}}
\def\mx#1{{\tilde m}_{\chi^{0}_#1}}
\def\mc#1{{\tilde m}_{\chi^{+}_#1}}
\def\mwi{{\tilde m}_{\omega i}}
\def\mxi{{\tilde m}_{\chi^{0}_i}}
\def\mci{{\tilde m}_{\chi^{+}_i}}
\def\mz{M_z}
\def\sw{\sin\theta_W}
\def\cw{\cos\theta_W}
\def\cb{\cos\beta}
\def\sb{\sin\beta}
\def\rwi{r_{\omega i}}
\def\rxj{r_{\chi j}}
\def\rfp{r_f'}
\def\Kik{K_{ik}}
\def\Fq2{F_{2}(q^2)}
\def\f{\({\cal F}\)}
\def\d1{{\f(\tilde c;\tilde s;\tilde W)+ \f(\tilde c;\tilde \mu;\tilde W)}}
\def\tw{\tan\theta_W}
\def\sec2w{sec^2\theta_W}
\def\ch{{\tilde\chi^{+}_1}}
\def\c2{{\tilde\chi^{+}_2}}

\def\tt{{\tilde\theta}}

\def\tp{{\tilde\phi}}

\def\mz{M_z}
\def\sw{\sin\theta_W}
\def\cw{\cos\theta_W}
\def\cb{\cos\beta}
\def\sb{\sin\beta}
\def\rwi{r_{\omega i}}
\def\rxj{r_{\chi j}}
\def\rfp{r_f'}
\def\Kik{K_{ik}}
\def\Fq2{F_{2}(q^2)}
\def\f{\({\cal F}\)}
\def\d1{{\f(\tilde c;\tilde s;\tilde W)+ \f(\tilde c;\tilde \mu;\tilde W)}}
\def\tw{\tan\theta_W}
\def\sec2w{sec^2\theta_W}

\section{Introduction}
Recently it was proposed that  mirror matter with masses in the several hundred GeV-TeV  range 
could exist and can  be  explored at the LHC energies
\cite{Ibrahim:2008gg}.  
Indeed many unified  models  do contain 
mirrors \cite{Georgi:1979md,Wilczek:1981iz,Senjanovic:1984rw,Babu:2002ti,Babu:2006rp}
and some  work on model building using mirrors can be found in \cite{Bars:1980mb,Adler:2002yg,Chavez:2006he}.
Also implications of mirrors are explored in several 
works \cite{Nandi:1981qr,Langacker:1988ur,Choudhury:2001hs,Csikor:1994jg,Montvay:1997zq,Triantaphyllou:1999uh}.
Specifically we discuss here the implications of mirror particles arising in vector-like combinations, e.g.,
in $5+\bar 5$ and $10+\bar 10$ multiplets in $SU(5)$ and in $16+\overline{16}$ multiplets in $SO(10)$.  
We assume that such vector-like multiplets escape gaining  superheavy masses but acquire vector-like
 masses of electroweak size (assumed to be much larger than the chiral masses they may acquire via 
 Yukawa couplings  to the MSSM Higgs doublets)
 which could lie in the several hundred GeV to TeV range or above. 
(The phenomenology of vector-like mutiplets has been discussed in several 
works \cite{Babu:2004xg,Babu:2008ge,Barger:2006fm,Lavoura:1992qd,Maekawa:1995ha,Babu:2004xg,sm}).
Such vector-like multiplets are consistent with the precision electroweak data specifically on the so called
S, T, U electroweak parameters and can be made consistent with the gauge coupling unification. 
Thus these models escape the constraints of  sequential generations   
\cite{Kribs:2007nz,Hung:2007ak,Holdom,Dubicki:2003am,Novikov:2002tk,Murdock:2008rx,Cakir:2008su,Cakir:2009ib,Liu:2009cc,Hewett:1986uu,Arnowitt:1987xk,Barger:1989dk,Frampton:1999xi}.
We allow a small mixing between the three sequential generations and the vector-like multiplets.
We  assume  these mixings to be  rather tiny so the effect of mixings with the normal matter in
the vector  multiplets will have  negligible effects on the analysis below and thus we focus on
the mixings of the mirrors with the ordinary three generations. Here again we assume that the 
mixings  are with the third generation.  
The reason for the suppression of mixings between the mirrors and the first two generations is 
  because the mixings generate a $V+A$ 
interactions for the ordinary generations. Now for the first two generations the $V-A$ structure
of the interactions is established to a great accuracy. However, this is less so for the case of the
third generation. \\

 Thus a small amount of mixing of the mirror generation and of the third generation
could be allowed but such mixings would be highly suppressed for the case of the first two
generations. Specifically analysis of experimental data to fix  the $\tau$ interactions allow 
for the possibility of a small $V+A$ type interaction \cite{Singh:1987hn,Dova:1998uj}.
A similar situation holds for the case of the third generation
 quarks \cite{Jezabek:1994zv,Nelson:1997xd,Abazov:2007ve}. 
This is also what one expects from an examination of the CKM matrix elements. 
For example, the mixings between the first and the second generations in the CKM matrix element
can be estimated by 
$V_{us}= \sqrt{m_d/m_s}$ which numerically is about 0.2. On the other hand the mixing between the
first generation and the third generation is given by $V_{ub}=\sqrt{m_d/m_b}$
 which numerically is 0.03 and thus much smaller. If we extrapolate these relations to include
 mixings with the vector-like multiplets 
  either mirror or non-mirror, then we can expect the mixing to be 
 roughly given by $V_{u B}=\sqrt{m_d/m_{V}}$. If we assume that $m_{V}=500$ GeV, the one has
 $V_{uB}\simeq 0. 003$  which is rather tiny.  In the leptonic sector such mixings will be even
 smaller. Thus in the leptonic sector the mixings between the first generation and the 
 vector multiplets  may be characterized by 
 $\sqrt{m_e/m_{V}}$ which  is 0.0009 for $m_e\simeq .5$ MeV and $m_V=500$ GeV.  
In the following as a simple approximation we will assume only mixings between the mirrors
 and the third generation and ignore mixings of the mirrors with the 
first two generations.  We mention here that the terminology mirror  has also appeared in the literature
in the context of mirror worlds \cite{okun,Mohapatra:2005ng} which is entirely different from 
the analysis here since in these models  one has  mirror matter with their own mirror  gauge group. 
There is  no relationship of the analysis here with those theories.\\

\section{Mirror mixings with MSSM particles}
 The superpotential of the model
for the lepton part, describing the mixings of the mirrors with the third generation leptons 
 may be written 
in the form
\beqn
W= \epsilon_{ij}  [f_{1} \hat H_1^{i} \hat \psi_L ^{j}\hat \tau^c_L
 +f_{1}' \hat H_2^{j} \hat \psi_L ^{i} \hat \nu^c_L\nonumber\\
+f_{2} \hat H_1^{i} \hat \chi^c{^{j}}\hat N_{L}
 +f_{2}' \hat H_2^{j} \hat \chi^c{^{i}} \hat E_{\tau L}]
+ f_{3} \epsilon_{ij}  \hat \chi^c{^{i}}\hat\psi_L^{j}\nonumber\\
 + f_{4} \hat \tau^c_L \hat  E_{\tau L}  +  f_{5} \hat \nu^c_L \hat N_{L}.
\label{2.1}
\eeqn
After spontaneous breaking of the electroweak symmetry, ($<H_1^1>=v_1/\sqrt{2} $ and $<H_2^2>=v_2/\sqrt{2}$),
we have the following set of mass terms written in 4-spinors for the fermionic sector

\beqn
-{\cal L}_m = \br\bar \tau_R ~ \bar E_{\tau R} \er
 \br
  f_1 v_1/\sqrt{2} ~ f_4\\
 f_3 ~ f_2' v_2/\sqrt{2}\er
 \br \tau_L\\
 E_{\tau L}\er\nonumber\\
  + \br\bar \nu_R ~ \bar N_R\er
 \br f'_1 v_2/\sqrt{2} ~ f_5\\
 -f_3 ~ f_2 v_1/\sqrt{2}\er \br \nu_L\\
 N_L\er  + H.c.\nonumber
 \label{2.2}
\eeqn 
Here 
the mass matrices are not  Hermitian and one needs
to use bi-unitary transformations to diagonalize them. Thus we write the linear transformations

\beqn
 \br\tau_R\\ 
 E_{\tau R}\er=D^{\tau}_R \br\tau_{1_R}\\
 E_{\tau 2_R} \er,\nonumber\\
\br \tau_L\\
 E_{\tau L}\er=D^{\tau}_L \br \tau_{1_L}\\
 E_{\tau 2_L}\er,
 \label{2.3}
\eeqn
such that
\beq
D^{\tau \dagger}_R \br f_1 v_1/\sqrt{2} ~ f_4\\
 f_3 ~ f_2' v_2/\sqrt{2}\er D^{\tau}_L=diag(m_{\tau_1},m_{\tau_2}).
\label{2.4}
\eeq

The same holds for the neutrino mass matrix 
\beqn
D^{\nu \dagger}_R \br f'_1 v_2/\sqrt{2} ~ f_5\\
 -f_3 ~ f_2v_1/\sqrt{2}\er D^{\nu}_L=diag(m_{\nu_1},m_{\nu_2}).
\label{2.5}
\eeqn

Here $\tau_1, \tau_2$ are the mass eigenstates and we identify the tau lepton 
with the eigenstate 1, i.e.,  $\tau=\tau_1$, and identify $\tau_2$ with a heavy 
mirror eigenstate  with a mass in the hundreds  of GeV. Similarly 
$\nu_1, \nu_2$ are the mass eigenstates for the neutrinos, 
where we identify $\nu_1$ with the light neutrino state and $\nu_2$ with the 
heavier mass eigenstate.
By multiplying Eq.(\ref{2.4}) by $D^{\tau \dagger}_L$ from the right and by
$D^{\tau}_R$ from the left and by multiplying Eq.(\ref{2.5}) by $D^{\nu \dagger}_L$
from the right and by $D^{\nu}_R$ from the left, one can equate the values of the parameter
$f_3$ in both equations and we can get the following relation
between the diagonlizing matrices $D^{\tau}$ and $D^{\nu}$
\beqn
m_{\tau 1} D^{\tau}_{R 21} D^{\tau *}_{L 11} +m_{\tau 2} D^{\tau}_{R 22} D^{\tau *}_{L 12}=\nonumber\\
-[m_{\nu 1} D^{\nu}_{R 21} D^{\nu *}_{L 11} +m_{\nu 2} D^{\nu}_{R 22} D^{\nu *}_{L 12}].
\label{2.6}
\eeqn
 Eq.(\ref{2.6}) is an important relation as it constraints the symmetry breaking parameters
 and this constraint must be taken into account in numerical analyses.
 
Let us now write the charged current interaction in the leptonic sector for the 3rd generation 
and for the mirror sector with the W boson. 
\beqn
{\cal{L}}_{CC}= -\frac{g_2}{2\sqrt 2} W^{\dagger}_{\mu}
[ \bar \nu\gamma^{\mu} (1-\gamma_5) \tau\nonumber\\
+\bar N\gamma^{\mu} (1+\gamma_5) E_{\tau}] + H.c.
\label{2.7}
\eeqn
In the mass diagonal basis the charged current interactions are given by 
\beqn
{\cal{L}}_{CC}=-\frac{g_2}{2\sqrt 2} W^{\dagger}_{\mu}
\sum_{\alpha,\beta,\gamma,\delta} \bar \nu_{\alpha} \gamma^{\mu} 
[D^{\nu\dagger}_{L\alpha\gamma} g_{\gamma\delta}^L D^{\tau}_{L\delta\beta}\nonumber\\ (1-\gamma_5)
+ D^{\nu\dagger}_{R\alpha\gamma} g_{\gamma\delta}^R D^{\tau}_{R\delta\beta} (1+\gamma_5)] 
 \tau_{\beta} +H.c. 
\label{2.8}
\eeqn
where $g^{L,R}_{\alpha\beta}$ are defined so that 
\beqn
g^L_{11}=1, g^L_{12}=0= g^L_{21}=g^L_{22}, \nonumber\\
g^R_{11}=0= g^R_{12}= g^R_{21}, g^R_{22}=1.
\label{2.9}
\eeqn
Assuming $D_L^{\tau}= D_R^{\tau}=D^{\tau}$ 
we may parametrize $D^{\tau}$  as follows
\beqn
D^{\tau}=
 {\left(
\begin{array}{cc}
\cos\theta & \sin\theta \cr
             -\sin\theta & \cos\theta
\end{array}\right)}. 
\label{2.91}
\eeqn
Similarly assuming  $D_L^{\nu}= D_R^{\nu}=D^{\nu}$  we may parametrize
the mixing between $\nu$ and $N$ by the angle $\phi$ where
\beqn
D^{\nu}=
 {\left(
\begin{array}{cc}
\cos\phi & \sin\phi \cr
             -\sin\phi & \cos\phi
 \end{array}\right)}.
 \label{2.92}
 \eeqn
With the above simplifications 
the charged current interaction including mirrors is given by \cite{Maalampi:1988va,mirrors,Ibrahim:2008gg}
\beqn
{\cal{L}}_{CC}= -\frac{g}{2\sqrt 2} W_{\mu}^{\dagger} J^{\mu}_C + H.c. 
\label{2.93}
\eeqn
where
\beqn
J^{\mu}_C= \{\bar \nu_1\gamma^{\mu} \tau_1 \cos(\theta-\phi)\nonumber\\ 
+\bar \nu_1 \gamma^{\mu} \tau_2 \sin(\theta-\phi) 
-\bar \nu_1\gamma^{\mu} \gamma_5 \tau_1 \cos(\theta+\phi)\nonumber\\ 
-\bar \nu_1 \gamma^{\mu} \gamma_5 \tau_2\sin(\theta+\phi)
-\bar \nu_2\gamma^{\mu} \tau_1 \sin(\theta-\phi) \nonumber\\
-\bar \nu_2 \gamma^{\mu}\gamma_5 \tau_1 \sin(\theta+\phi) 
+\bar \nu_2\gamma^{\mu}  \tau_2 \cos(\theta-\phi) \nonumber\\
+\bar \nu_2 \gamma^{\mu} \gamma_5 \tau_2\cos(\theta+\phi)\}. 
\label{2.10}
\eeqn
Here $\tau_1, \tau_2$ are the mass eigenstates for the charged leptons, with 
$\tau_1$ identified  as the physical tau state, 
and $\nu_1, \nu_2$ are the mass eigenstates for the neutrino
 with $\nu_1$ identified as the observed 
neutrino. [The result of Ref. \cite{Ibrahim:2008gg} regarding 
 Eq.(\ref{2.93}) and  Eq.(\ref{2.10}) agrees with  Eq.(1) 
of \cite{mirrors} after correction of a typo in  \cite{mirrors}].
The result of  Eq.(\ref{2.93}) and Eq.(\ref{2.10})  correctly reduces  to the result of Eq.(\ref{2.7}) in
the limit when $\theta=0=\phi$. \\

 Next we  consider  the mixings of the charged sleptons and the charged mirror sleptons. 
 These mixings will in general contain new sources of CP phases beyond those in the
  MSSM (for a review see \cite{Ibrahim:2007fb}). However, for the analysis here we set these
 phases to zero.
The mass matrix in the basis $(\tilde  \tau_L, \tilde E_L, \tilde \tau_R, 
\tilde E_R)$  will be a  hermitian $4\times 4$ mass matrix.  
  We diagonalize this hermitian mass$^2$ matrix  by the
following unitary transformation 
\beqn
 \tilde D^{\tau \dagger} M^2_{\tilde \tau} \tilde D^{\tau} = diag (M^2_{\tilde \tau_1},  
M^2_{\tilde \tau_2}, M^2_{\tilde \tau_3},  M^2_{\tilde \tau_4}).
\label{2.11}
\eeqn
The mixing between $\tilde  \tau_L$ and $ \tilde E_R$ vanishes as well as the
mixing between $\tilde  \tau_R$ and $ \tilde E_L$. The mixing of $\tilde  \tau_L$ and $ \tilde \tau_R$
is given by $m_{\tau}(A_{\tau}-\mu \tan \beta)$ and of $\tilde  E_L$ and $ \tilde E_R$ is given
by $m_{E}(A_{E}-\mu \cot \beta)$. We could choose the parameters of the superpotential and the
soft breaking parameters such that the latter two mixings vanish as well. Thus
We make the assumption that the mixings between the mirror charged sleptons
and the staus arising principally from the mixing between  $\tilde  \tau_L$ and $ \tilde E_L$
and between $\tilde \tau_R$ and  $\tilde E_R)$.  Under  this assumption one has
\beqn
\tilde{D}^{\tau}={\left(
\begin{array}{cccc}
\cos\tilde{\theta}_1 & \sin\tilde{\theta}_1 &0&0 \cr
       -\sin\tilde{\theta}_1 & \cos\tilde{\theta}_1 &0&0\cr
0&0&\cos\tilde{\theta}_2 & \sin\tilde{\theta}_2\cr
0&0&-\sin\tilde{\theta}_2 & \cos\tilde{\theta}_2
 \end{array}\right)}.\nonumber
 \label{2.12}
\eeqn
A similar mass$^2$ matrix exists in the sneutrino sector.
In the basis $(\tilde  \nu_L, \tilde N_L, \tilde \nu_R, 
\tilde N_R)$  we again get a hermitian $4\times 4$ matrix
which can be diagonalized  by the
following unitary transformation 
\beqn
 \tilde D^{\nu\dagger} M^2_{\tilde \nu} \tilde D^{\nu} = diag (M^2_{\tilde \nu_1},  
M^2_{\tilde \nu_2}, M^2_{\tilde \nu_3},  M^2_{\tilde \nu_4}).
\label{2.13}
\eeqn
The physical tau and neutrino states are $\tau\equiv \tau_1, \nu\equiv \nu_1$,
and the states $\tau_2, \nu_2$ are heavy states with mostly mirror particle content. 
The states $\tilde \tau_i, \tilde \nu_i; ~i=1-4$ are the slepton and sneutrino states. 
As in the case of the mixings of the staus and of the charged mirror  sleptons we
assume that the mixings of sneutrinos  and of the sneutrinos arises principally 
due to mixings between $\tilde  \nu_L$ and $\tilde N_L$   and between 
$\tilde \nu_R$ and  $\tilde N_R$.  Under these simplifying assumptions one has
\beqn
\tilde{D}^{\nu}={\left(
\begin{array}{cccc}
\cos\tilde{\phi}_1 & \sin\tilde{\phi}_1 &0&0 \cr
             -\sin\tilde{\phi}_1 & \cos\tilde{\phi}_1 &0&0\cr
0&0&\cos\tilde{\phi}_2 & \sin\tilde{\phi}_2\cr
0&0&-\sin\tilde{\phi}_2 & \cos\tilde{\phi}_2
 \end{array}\right)}.\nonumber
 \label{2.14}
\eeqn
Under further simplifying assumptions one can get $\tilde \theta_1=\tilde \theta_2=\tilde \theta$,
$\tilde \phi_1=\tilde \phi_2=\tilde \phi$.
For the case of  
no mixing these limit as  follows 
\beqn
\tilde \tau_1\to \tilde \tau_L, ~\tilde \tau_2\to \tilde E_L, ~\tilde \tau_3\to \tilde \tau_R, ~
\tilde \tau_4\to \tilde E_R,\nonumber\\
\tilde \nu_1\to \tilde \nu_L, ~\tilde \nu_2\to \tilde N_L, ~\tilde \nu_3\to \tilde \nu_R, ~
\tilde \nu_4\to \tilde N_R.
\label{2.15}
\eeqn

Next we look at the neutral current interactions and focus on the charged
	leptons. Here the Z  boson interactions are given by

	\beqn
	{\cal{L}}_{NC}= -\frac{g}{4 \cos\theta_W} Z_{\mu}[ \bar \tau\gamma^{\mu} (4x-1+\gamma_5)\tau\nonumber\\
	+  \bar E_{\tau}\gamma^{\mu} (4x-1-\gamma_5)E_{\tau}],
	\label{nc}
	\label{2.16}
	\eeqn
where $x=\sin^2\theta_W$.  We can also write these results in the mass diagonal basis. 

\beqn
{\cal{L}}_{NC}= -\frac{g}{4\cos\theta_W}Z_{\mu}J^{\mu}_N,\nonumber\\
J^{\mu}_N=
\{\bar \tau_1\gamma^{\mu} (4\cos^2\theta_W -1+\cos 2\theta \gamma_5)\tau_1\nonumber\\ 
+\bar \tau_2\gamma^{\mu} (4\cos^2\theta_W -1-\cos 2\theta \gamma_5)\tau_2\nonumber\\
+\bar \tau_1\gamma^{\mu}\gamma_5 \sin 2\theta \tau_2+
\bar \tau_2\gamma^{\mu}\gamma_5 \sin 2\theta \tau_1\},
\label{2.17}
\eeqn
We discuss next the implications of the above interactions for the anomalous magnetic moment of 
the tau lepton and for the magnetic moment of the tau neutrino. 

\section{Tau neutrino magnetic moment\label{magnetic}} 
   The existence of neutrino masses is now well established with the current experimental limits
   on their masse arising from WMAP obeying the constraint \cite{Spergel:2006hy}
   \beqn
   \sum_i |m_{\nu_i}|\le (.7-1) {\rm eV}.
   \label{3.1}
  \eeqn 
The nature of the neutrino masses is not yet determined. They could be either Majorana
or Dirac. The smallness of the neutrino masses is easy to understand if the masses of 
the neutrinos are Majorana as they could arise from a see saw mechanism. On other 
it is not so easy to understand their smallness if the masses are Dirac and this 
topic continues to be subject of much current activity. If the neutrino masses are Dirac, they
can have both a magnetic moment and an electric dipole moment.  We will focus here 
on the magnetic moment. In the standard model enhanced by a right handed  neutrino,
a tau neutrino has  a magnetic moment arising from the exchange of a W boson
which gives  a magnetic moment of \cite{CabralRosetti:1999ad,Dvornikov:2003js}
\beqn
\mu_{\nu}=O(10^{-19}) (m_{\nu}/eV)\mu_B.
\label{3.2}
\eeqn
where $\mu_B=(e/2m_e)$ is the Bohr magneton.   One may compare this result with the current limits 
on the tau neutrino   magnetic moment from experiment which gives \cite{exp3}
 \beqn
  |\mu(\nu_{\tau})|\leq 1.3\times 10^{-7} \mu_B, 
  \label{3.3}
 \eeqn
 One may note that the SM prediction is twelve orders of magnitude smaller than the experimental 
 limit, and thus beyond the reach of experimental test in any near future experiment. 
   One can extend the above analysis to MSSM and include the supersymmetric 
   exchange, i.e., the chargino exchange. However, inclusion of the supersymmetric exchanges
   does not change the order of the contribution which is still of the size given by Eq,(\ref{3.2}). 
   However, there is very drastic change when one includes the mirror particles  and their
   supersymmetric partners. Specifically, an extension to include mirrors brings in the following
   additional contributions: (1) Mirror lepton and W exchange, and (2) Mirror  slepton and 
   chargino exchange. In this case neutrino magnetic moments which are larger by  as much
   as ten orders of magnitude than the SM contribution can be obtained. Thus the 
   magnetic moments in these models come within the realms of observability if
   improvement in experiment by one to two orders of magnitude can occur. \\

     In addition to the above one must also take into account the constraints arising from
     the limits on the $\tau$ magnetic moment. The current experimental limit on the $\tau$
     magnetic moment is 
 \beqn
 a_{\tau}(SM)= 117721(50\times 10^{-8},
 \label{3.4}
 \eeqn
where $a_{\tau}= (g_{\tau}-2)/2$ while the current experimental limit on the parameter is 
\beqn
-0.052 < a_{\tau}(exp) < 0.013.
\label{3.5}
\eeqn
  A comparison of Eq.(\ref{3.4}) and Eq.(\ref{3.5}) shows that the 
  the current experimental sensitivity is just below an order of magnitude of what the SM predicts.
  Because of these constraints it is important to examine what the correction of the mirror states
  is to the SM correction. In Ref. \cite{Ibrahim:2008gg}
   an analysis of the $\tau$ anomalous magnetic moment 
  at one loop is given taking into account several contributions. Thus the analysis given in 
  Ref.\cite{Ibrahim:2008gg} includes the Standard Model contribution involving the W and Z exchange as well
  as the supersymmetric contribution involving the exchange of the chargino and the neutralinos.
  Additionally exchanges of mirror  leptons and mirror sleptons are included. The analysis
  shows that the contributions from the mirror sector this time are of the same order as the
  contribution from the SM sector. This in sharp contrast to the contribution to the $\tau$ neutrino
  case where the contributions from the mirror sector are orders of magnitude larger than
  the SM contribution.   
  In Table (1) we give an analysis of the neutrino magnetic moment and of the $\tau$ anomalous
  magnetic moment within the framework of the supergravity grand unified model \cite{msugra}. 
  \begin{figure}
 \vspace{-2cm}
      \scalebox{2.25}
      {
       \hspace{-1cm}      
       \includegraphics[width=5cm,height=6cm]{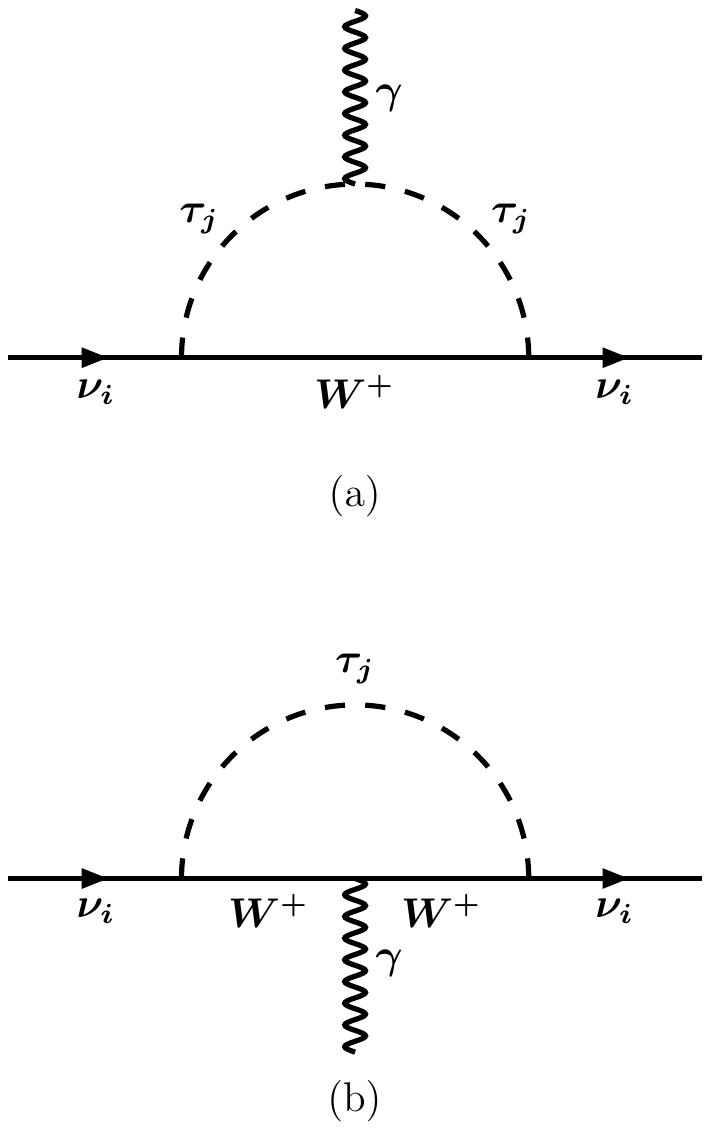}}
     \vspace{-7cm}
\caption{The loop contributions from the exchange of the leptons and mirror
leptons  to the magnetic 
dipole moment of neutrinos ($\nu_i$) and via the exchange of $W^+$ boson.
 There are similar loop contributions from the exchange of sleptons, mirror sleptons
 and from the exchange of charginos.}
\label{fig1}
       \end{figure}
  \begin{center} \begin{tabular}{|c|c|c|c|c|}
\multicolumn{5}{c}{Table~1:  } \\
\hline
$\theta$ & $\tilde{\theta}$  &$\tilde{\phi}$ & $\Delta a_{\tau}\times 10^6$
 & $\mu_{\nu}/\mu_B\times 10^{10}$\\
\hline
$0.2$     &  $0.4$  &  $0.3$     &   $6.95 $    &    $-9$\\
 \hline
$0.15$     &  $0.45$  &  $0.35$     &   $3.46$    &    $-5.3 $ \\ 
\hline
 $0.10$     &  $0.3$  &  $0.2$     &   $1.64$    &    $-2.4 $ \\
 \hline
\end{tabular}\\~\\
\label{tab:1}
\noindent
\end{center}
Table caption:  A sample illustration of the 
contributions to the magnetic moments of  $\nu_{\tau}$
 and of $\tau$  
including  corrections from exchange of the third generation leptons and their 
superpartners and  from exchange of the mirror particles and mirror super partners.
The analysis is for the parameter set 
$m_0=400$, $m_{1/2}=150$, $\tan\beta=20$, $A_0=400$,  $\mu>0$,
 $m_E=200$,  $m_N=220$,
 $M_{\tilde{\tau}_{11}}=400$,
$M_{\tilde{\tau}_{22}}=500$, $m_{\tilde{\nu}_{11}}=420$ and
 $m_{\tilde{\nu}_{22}}=520$. The mixing angles $\theta$ , $\tilde \theta$
 and $\tilde \phi$ are as exhibited.
All masses are in units of GeV and all angles are in radian.

\section{Signatures for mirrors at the LHC\label{sig}}
A variety of signatures can arise for the mirror fermions and mirror sfermions if they exist
at the LHC. We discuss these below.

\subsection{Lepton and jet signatures of mirrors at the LHC}
 In the extension of MSSM with a mirror generation,  one has new fermions,
 i.e., the mirror fermions $B, T, E, N$  (see the Appendix for notation) all Dirac, 
 as well as mirror sfermions.  
 For the mirror  sleptons  we have the  new states 
 \beqn
\tilde E_1, \tilde E_2, 
~\tilde N_{1}, ~\tilde N_{2},
~\tilde N_{3},  
\label{4.8}
\eeqn
where $\tilde E_1, \tilde E_2$ are the charged mirror sleptons and 
$\tilde N_1, \tilde N_2, \tilde N_3$ are 
the three new sneutrino states. The reason of these extra three sneutrino is
that we started out with two extra chiral singlets, one in the MSSM sector and
another in the mirror sector. Along with the two chiral neutrino states that arise from the doublet
they produce four sneutrino states, one of which is in the MSSM sector and the the other
three are new and listed in the equation above.
 For the mirror squarks we have 
\beqn
 \tilde B_1, \tilde B_2, \tilde T_1, \tilde T_2.
 \label{4.9} 
\eeqn
The decay modes of the mirrors will produce interesting signatures. Thus, for example,
the decay of the mirror lepton $E^-$ gives
\beqn
E^-\to \tau^- Z\to \tau^- l^+l^-, \tau^- {\rm jets}, 
\label{4.10}
\eeqn
and decay of the Dirac $N$ gives
\beqn
N\to E^-W^+, ~N\to \nu_{\tau}\gamma.
\label{4.11}
\eeqn
where the decay $N\to \nu_{\tau}\gamma$ occurs via transition magnetic moment.
Similarly there will be the mirror slepton with the decay 
\beqn
\tilde E^-\to \tilde \tau^- Z, \tau^-\tilde \chi_i^0 \to \tau^- l^+l^-+E_T^{\rm miss},
\label{4.12}
\eeqn
where $E_T^{\rm miss}$ is typically the lightest neutralino (the LSP).
Similarly the mirror sneutrinos will have the decay
\beqn
\tilde N_i \to \tilde E^- W^+, E^- \tilde \chi^+  \to \tau^-l^+ + E_T^{\rm miss}.
\label{4.13}
\eeqn
The above will give rise to the  following processes at the LHC.  
\beqn 
pp\to W^*\to EN
\to  [\tau l_i\bar l_i,  3\tau, \tau +2 {\rm jets}]  E_T^{\rm miss}\nonumber\\
p p \to Z^* \to E^+E^-  \to \nonumber\\2\tau 4l, 4\tau 2l, 6\tau, 2\tau 2l 2{\rm jets}, 2 \tau 4{\rm jets}.
\label{4.14}
\eeqn
where $l_1,l_2= e, \mu$. Similar signatures but with much more missing energy arise via
the production and decay of mirror sfermions. 
We note that the trileptonic signature of Eq.(\ref{4.14}) is distinctly different from the trileptonic 
signature arising from the off shell decay of a $W$ in supersymmetry \cite{Nath:1987sw}.
Similarly the 4 lepton decay in Eq.(\ref{4.14}) is also distinctly different from the  4 leptonic
decay in supersymmetry \cite{ti}. This is because in Eq.(\ref{4.14}) there is always a tau in the final
final state. 
\subsection{Forward-backward asymmetry}
The forward backward asymmetry defined by 
$A_{FB} = \left(\int_0^1 (d\sigma/dz)dz -\int_{-1}^0(d\sigma/dz)dz\right)$ $/\int_{-1}^1(d\sigma/dz)dz$
in the process $f\bar f \to f_4\bar f_4$ vs the process
$f\bar f \to f_m \bar f_m$ where $f_4$ is the 4th generation fermion and $f_m$ is the 
mirror fermion can allow one to discriminate between ordinary fermions  and  mirror fermions
 since $A_{FB}$ is sensitive to the Lorentz structure of the interaction. Thus a 
measurement of the forward -backward asymmetry provides an important test of the presence
of a $V+A$ vs a $V-A$ interaction and will allow one to discriminate   mirrors  from normal particles 
in collider experiments.  Finally we mention that simulations of models with vector-like 
multiplets  are 
needed in order to make concrete predictions of the event rates of various signatures listed in
this section for LHC luminosities  similar to the ones done for the supergravity models 
(for some recent works see \cite{Feldman:2007zn}).
\subsection{FCNC processes}
The presence of mirror particles  in  vector-like
multiplets which mix with the third generation would tend to vitiate the GIM mechanism 
and produce FCNC processes. This will lead to couplings of the Z boson of the
type  $Z\bar \tau E$, $Z \bar b B$, $Z\bar t T$. Thus in Drell-Yan processes one
will be able to produce  final states with $\bar \tau E$, $\bar b B$ and $\bar t T$ etc.
Of course such processes would be suppressed by small mixing angles  as well
as by the largeness of the mirror fermion masses. One needs  detailed simulations
on the production of such events  at the LHC energies and luminosities.

\section{Conclusion\label{conclusion}}
We have discussed here  the possibility that mirror particles may exist with low masses,
i.e., masses in the electroweak region. The simplest possibility is that the mirrors are part
of  vector like multiplets which survive superheavy mass growth but gain masses of electroweak
size.  
  Further, mixings can  arise between the vector-like  multiplets 
   and the three sequential generations.
 While such mixings are severely limited by experimental data for the first two generations,
 a small mixing between the vector-like multiplets and the third generation is not excluded.
We focus here on the mixings between the mirrors and the third generation.
As a consequence of this mixing the third generation fermions develop a small $(V+A)$ interaction
in addition to their normal $(V-A)$ interactions.  One consequence of this new interaction is 
an enhancement of the $\tau$ neutrino magnetic moment by as much as ten orders of 
magnitude bringing it within the realm of observation in future experiment.  There are 
also important collider implications of mirror fermions and mirror sfermions.
Is it estimated  that mirror particles can be observed
at the LHC with about $50$fb$^{-1}$ of data.  Further, there would be some very unique
features associated with the production of the mirror particles. Specifically because of the
mixing of the mirrors with the third generation, the leptonic decay modes
of the mirrors will always contain a tau lepton. Other signatures involve the forward-backward
asymmetry in the production of the mirror fermion -anti mirror fermion pair  which is different
from that of an ordinary fermion-anti fermion pair.

\noindent
{\large\bf Acknowledgments}\\ 
This research is  supported in part by NSF grant PHY-0757959. 
\section{Appendix: Mirror fermions}
  The transformation properties of the mirror fermions  and  those of the  ordinary
fermions are as below. 
 Thus the ordinary  quarks  under $SU(3)_C\times SU(2)_L \times U(1)_Y$  transform 
 as follows
\beqn
q^T\equiv \left(t_L, b_L
\right)
\sim(3,2,\frac{1}{6}),\nonumber\\ t^c_L\sim (3^*,1,-\frac{2}{3}),
~b^c_L\sim (3^*,1,\frac{1}{3}).
\eeqn
The corresponding transformations for the mirror quarks are
\beqn
{Q}^{cT} \equiv \left(B^c_L, T^c_L
\right)
\sim(3^*,2,-\frac{1}{6}),\nonumber\\
 T_L\sim (3,1,\frac{2}{3}),
~ B_L\sim (3^*,1, -\frac{1}{3}).
\eeqn
Analogously,  the transformation properties of the ordinary leptons  under $SU(3)_C\times SU(2)_L \times U(1)_Y$  
are 
\beqn
\psi_L^T\equiv \left(\nu_L, \tau_L
\right) \sim(1,2,- \frac{1}{2}),\nonumber\\  \tau^c_L\sim (1,1,1),
 ~\nu^c_L\sim (1,1,0),
\eeqn
while those of the mirror leptons are 
\beqn
\chi^{cT}\equiv \left(E_{\tau L}^c, N_L^c
\right)
\sim(1,2,\frac{1}{2}),\nonumber\\
 E_{\tau L}\sim (1,1,-1), ~N_L\sim (1,1,0).
\eeqn


\begin{thebibliography}{999}

\bibitem{Ibrahim:2008gg}
  T.~Ibrahim and P.~Nath,
  Phys.\ Rev.\  D {\bf 78}, 075013 (2008)
  [arXiv:0806.3880 [hep-ph]].

\bibitem{Georgi:1979md}
  H.~Georgi,
  Nucl.\ Phys.\  B {\bf 156}, 126 (1979).

\bibitem{Wilczek:1981iz}
  F.~Wilczek and A.~Zee,
  Phys.\ Rev.\  D {\bf 25}, 553 (1982).

\bibitem{Senjanovic:1984rw}
  G.~Senjanovic, F.~Wilczek and A.~Zee,
  Phys.\ Lett.\  B {\bf 141}, 389 (1984).

\bibitem{Babu:2002ti}
  K.~S.~Babu, S.~M.~Barr and B.~s.~Kyae,
  Phys.\ Rev.\  D {\bf 65}, 115008 (2002)
  [arXiv:hep-ph/0202178].

\bibitem{Babu:2006rp}
  K.~S.~Babu, I.~Gogoladze, P.~Nath and R.~M.~Syed,
  Phys.\ Rev.\  D {\bf 74}, 075004 (2006)
  [arXiv:hep-ph/0607244];
 K.~S.~Babu, I.~Gogoladze, P.~Nath and R.~M.~Syed,
  Phys.\ Rev.\  D {\bf 74}, 075004 (2006)
  [arXiv:hep-ph/0607244];
 P.~Nath and R.~M.~Syed,
  arXiv:0909.2380 [hep-ph].


\bibitem{Bars:1980mb}
  I.~Bars and M.~Gunaydin,
  Phys.\ Rev.\ Lett.\  {\bf 45}, 859 (1980).

\bibitem{Adler:2002yg}
  S.~L.~Adler,
  Phys.\ Lett.\  B {\bf 533}, 121 (2002)
  [arXiv:hep-ph/0201009].

\bibitem{Chavez:2006he}
  H.~Chavez and J.~A.~Martins Simoes,
  Nucl.\ Phys.\  B {\bf 783}, 76 (2007)
  [arXiv:hep-ph/0610231];
P. Csikor and Z. Fodor, hep-ph/9205222;
 V.~Elias and S.~Rajpoot,
  Phys.\ Lett.\  B {\bf 134}, 201 (1984);
 S.~Rajpoot and J.~G.~Taylor,
  Phys.\ Lett.\  B {\bf 142}, 365 (1984).

\bibitem{Nandi:1981qr}
  S.~Nandi, A.~Stern and E.~C.~G.~Sudarshan,
  Phys.\ Rev.\  D {\bf 26}, 2522 (1982).

\bibitem{Langacker:1988ur}
  P.~Langacker and D.~London,
  Phys.\ Rev.\  D {\bf 38}, 886 (1988).

\bibitem{Choudhury:2001hs}
  D.~Choudhury, T.~M.~P.~Tait and C.~E.~M.~Wagner,
  Phys.\ Rev.\  D {\bf 65}, 053002 (2002)
  [arXiv:hep-ph/0109097].

\bibitem{Csikor:1994jg}
  F.~Csikor and I.~Montvay,
  Phys.\ Lett.\  B {\bf 324}, 412 (1994)
  [arXiv:hep-ph/9401290].

\bibitem{Montvay:1997zq}
  I.~Montvay,
  arXiv:hep-ph/9708269.

\bibitem{Triantaphyllou:1999uh}
  G.~Triantaphyllou,
  Int.\ J.\ Mod.\ Phys.\  A {\bf 15}, 265 (2000)
  [arXiv:hep-ph/9906283].

\bibitem{Babu:2008ge}
  K.~S.~Babu, I.~Gogoladze, M.~U.~Rehman and Q.~Shafi,
  Phys.\ Rev.\  D {\bf 78}, 055017 (2008)
  [arXiv:0807.3055 [hep-ph]].

\bibitem{Babu:2004xg}
  K.~S.~Babu, I.~Gogoladze and C.~Kolda,
  arXiv:hep-ph/0410085.



\bibitem{Barger:2006fm}
  V.~Barger, J.~Jiang, P.~Langacker and T.~Li,
  Int.\ J.\ Mod.\ Phys.\  A {\bf 22}, 6203 (2007)
  [arXiv:hep-ph/0612206].

\bibitem{Lavoura:1992qd}
  L.~Lavoura and J.~P.~Silva,
  Phys.\ Rev.\  D {\bf 47}, 1117 (1993).

\bibitem{Maekawa:1995ha}
  N.~Maekawa,
  Phys.\ Rev.\  D {\bf 52}, 1684 (1995).

\bibitem{Babu:2004xg}
  K.~S.~Babu, I.~Gogoladze and C.~Kolda,
  arXiv:hep-ph/0410085.

\bibitem{sm}
S. Martin, Talk at SUSY09. 
$http://nuweb.neu.edu/susy09/talks/Talk_742-Martin.pdf$

\bibitem{Kribs:2007nz}
  G.~D.~Kribs, T.~Plehn, M.~Spannowsky and T.~M.~P.~Tait,
  Phys.\ Rev.\  D {\bf 76}, 075016 (2007)
  [arXiv:0706.3718 [hep-ph]];
   R.~Fok and G.~D.~Kribs,
  arXiv:0803.4207 [hep-ph].

\bibitem{Hung:2007ak}
  P.~Q.~Hung and M.~Sher,
  Phys.\ Rev.\  D {\bf 77}, 037302 (2008)
  [arXiv:0711.4353 [hep-ph]].

\bibitem{Holdom}
  B.~Holdom,
  JHEP {\bf 0608}, 076 (2006)
  [arXiv:hep-ph/0606146];
  JHEP {\bf 0703}, 063 (2007)
  [arXiv:hep-ph/0702037].

\bibitem{Dubicki:2003am}
  J.~E.~Dubicki and C.~D.~Froggatt,
  Phys.\ Lett.\  B {\bf 567}, 46 (2003)
  [arXiv:hep-ph/0305007].

\bibitem{Novikov:2002tk}
  V.~A.~Novikov, L.~B.~Okun, A.~N.~Rozanov and M.~I.~Vysotsky,
  JETP Lett.\  {\bf 76}, 127 (2002)
  [Pisma Zh.\ Eksp.\ Teor.\ Fiz.\  {\bf 76}, 158 (2002)]
  [arXiv:hep-ph/0203132].

\bibitem{Murdock:2008rx}
  Z.~Murdock, S.~Nandi and Z.~Tavartkiladze,
  arXiv:0806.2064 [hep-ph].

\bibitem{Cakir:2008su}
  O.~Cakir, H.~Duran Yildiz, R.~Mehdiyev and I.~Turk Cakir,
  arXiv:0801.0236 [hep-ph].

\bibitem{Cakir:2009ib}
  I.~T.~Cakir, H.~D.~Yildiz, O.~Cakir and G.~Unel,
  arXiv:0908.0123 [hep-ph].

\bibitem{Liu:2009cc}
  C.~Liu,
  arXiv:0907.3011 [hep-ph].

\bibitem{Hewett:1986uu}
  J.~L.~Hewett and T.~G.~Rizzo,
  Phys.\ Rev.\  D {\bf 35}, 2194 (1987).

\bibitem{Arnowitt:1987xk}
  R.~L.~Arnowitt and P.~Nath,
  Phys.\ Rev.\  D {\bf 36} (1987) 3423.

\bibitem{Barger:1989dk}
  V.~D.~Barger, J.~L.~Hewett and T.~G.~Rizzo,
  Mod.\ Phys.\ Lett.\  A {\bf 5}, 743 (1990).

\bibitem{Frampton:1999xi}
  P.~H.~Frampton, P.~Q.~Hung and M.~Sher,
  Phys.\ Rept.\  {\bf 330}, 263 (2000)
  [arXiv:hep-ph/9903387].

\bibitem{Singh:1987hn}
  S.~Singh and N.~K.~Sharma,
  Phys.\ Rev.\  D {\bf 36}, 160 (1987);
  S.~Singh and N.~K.~Sharma,
  Phys.\ Rev.\  D {\bf 36}, 3387 (1987).

\bibitem{Dova:1998uj}
  M.~T.~Dova, J.~Swain and L.~Taylor,
  Nucl.\ Phys.\ Proc.\ Suppl.\  {\bf 76} (1999) 133
  [arXiv:hep-ph/9811209].

\bibitem{Jezabek:1994zv}
  M.~Jezabek and J.~H.~Kuhn,
  Phys.\ Lett.\  B {\bf 329}, 317 (1994)
  [arXiv:hep-ph/9403366].

\bibitem{Nelson:1997xd}
  C.~A.~Nelson, B.~T.~Kress, M.~Lopes and T.~P.~McCauley,
  Phys.\ Rev.\  D {\bf 56}, 5928 (1997)
  [arXiv:hep-ph/9707211];

\bibitem{Abazov:2007ve}
  V.~M.~Abazov {\it et al.}  [D0 Collaboration],
  Phys.\ Rev.\ Lett.\  {\bf 100}, 062004 (2008)
  [arXiv:0711.0032 [hep-ex]].
\bibitem{okun}
Yu. Kobzarev, L.B. Okun and I. Ya. Pomeranchuk, Sov. J.  of  Nucl. Phys. {\bf 3},
837 ( 1966).

\bibitem{Mohapatra:2005ng}
  R.~N.~Mohapatra, S.~Nasri and S.~Nussinov,
  Phys.\ Lett.\  B {\bf 627}, 124 (2005)
  [arXiv:hep-ph/0508109].



\bibitem{Maalampi:1988va}
  J.~Maalampi and M.~Roos,
  Phys.\ Rept.\  {\bf 186}, 53 (1990).

  \bibitem{mirrors}
J. Maalampi, J.T. Peltoniemi, and M. Roos, PLB 220, 441(1989).

\bibitem{Ibrahim:2007fb}
  T.~Ibrahim and P.~Nath,
  Rev.\ Mod.\ Phys.\  {\bf 80}, 577 (2008);
  arXiv:hep-ph/0210251.
  
\bibitem{Spergel:2006hy}
  D.~N.~Spergel {\it et al.}  [WMAP Collaboration],
  Astrophys.\ J.\ Suppl.\  {\bf 170} (2007) 377
  [arXiv:astro-ph/0603449].

\bibitem{CabralRosetti:1999ad}
  L.~G.~Cabral-Rosetti, J.~Bernabeu, J.~Vidal and A.~Zepeda,
  Eur.\ Phys.\ J.\  C {\bf 12}, 633 (2000)
  [arXiv:hep-ph/9907249].

\bibitem{Dvornikov:2003js}
  M.~Dvornikov and A.~Studenikin,
  Phys.\ Rev.\  D {\bf 69}, 073001 (2004)
  [arXiv:hep-ph/0305206].

  \bibitem{exp3}
  S. ~N. Gninenko,  Phys. Lett. B, {\bf 452},  414 (1999).

\bibitem{msugra}
A.~H.~Chamseddine, R.~Arnowitt and P.~Nath,
  Phys.\ Rev.\ Lett.\  {\bf 49} (1982) 970;

\bibitem{Nath:1987sw}
  P.~Nath and R.~Arnowitt,
  Mod.\ Phys.\ Lett.\  A {\bf 2}, 331 (1987);
   H.~Baer, C.~h.~Chen, F.~Paige and X.~Tata,
  Phys.\ Rev.\  D {\bf 50}, 4508 (1994);
 V.~D.~Barger, C.~Kao and T.~j.~Li,
  Phys.\ Lett.\  B {\bf 433}, 328 (1998).

  \bibitem{ti}
 T.~Ibrahim,
  Phys.\ Rev.\  D {\bf 77}, 065028 (2008)
  [arXiv:0803.4134 [hep-ph]].

\bibitem{Feldman:2007zn}
  D.~Feldman, Z.~Liu and P.~Nath,
  Phys.\ Rev.\ Lett.\  {\bf 99}, 251802 (2007);
 Phys.\ Lett.\  B {\bf 662}, 190 (2008);
  JHEP {\bf 0804}, 054 (2008).


\end{thebibliography}
\end{document}